%% file: main.tex
\documentclass[conference]{IEEEtran}
\IEEEoverridecommandlockouts
\usepackage{cite}
\usepackage{amsmath,amssymb,amsfonts}
\usepackage{graphicx}
\usepackage{textcomp}
\usepackage{xcolor}

\usepackage{hyperref}
\usepackage{xurl}
\usepackage{multirow}
\usepackage{makecell}
\usepackage{booktabs}
\usepackage{cleveref}
\usepackage{multicol}
\usepackage{tcolorbox}
\usepackage{algorithm}
\usepackage{algpseudocode}
\usepackage{tikz}
\usetikzlibrary{arrows.meta, positioning, shapes.geometric}

\usepackage{tikz}
\usetikzlibrary{arrows.meta,positioning,calc,fit}

\usepackage{listings}

\lstdefinestyle{promptstyle}{
  basicstyle=\scriptsize\ttfamily,
  breaklines=true,
  breakatwhitespace=false,
  columns=fullflexible,
  keepspaces=true,
  frame=single,
  framerule=0.4pt,
  rulecolor=\color{black!60},
  backgroundcolor=\color{gray!10},
  xleftmargin=2pt,
  xrightmargin=2pt
}

\def\BibTeX{{\rm B\kern-.05em{\sc i\kern-.025em b}\kern-.08em
    T\kern-.1667em\lower.7ex\hbox{E}\kern-.125emX}}
\begin{document}

\title{The Spoken Wikipedia Presentation Corpus\\

\thanks{This work was partially funded by the German Federal Ministry of Research, Technology and Space (BMFTR) under Grant 16DHBKI026.}
}

\author{
\IEEEauthorblockN{Thomas Ranzenberger, Steffen Freisinger, Tobias Bocklet, Korbinian Riedhammer}
\IEEEauthorblockA{
\textit{Technische Hochschule Nürnberg}\\
Nürnberg, Germany \\
\href{}{\{firstname.lastname\}@th-nuernberg.de}
}
}

\makeatletter
\def\ps@IEEEtitlepagestyle{%
  \def\@oddfoot{\mycopyrightnotice}%
  \def\@evenfoot{}%
}
\def\mycopyrightnotice{%
  \begin{minipage}{\textwidth}
  \centering \scriptsize
  Copyright~\copyright~2026 IEEE. Personal use of this material is permitted. Permission from IEEE must be obtained for all other uses, in any current or future media, including reprinting/republishing this material for advertising or promotional purposes, creating new collective works, for resale or redistribution to servers or lists, or reuse of any copyrighted component of this work in other works.
  \end{minipage}
}
\makeatother

\maketitle

\begin{abstract}
We present the Spoken Wikipedia Presentation Corpus, an extension of the Spoken Wikipedia Corpora featuring LLM-generated slide decks for multimodal ASR. Slides are created from LLM-segmented sections using a hybrid pipeline that combines LLM-based content planning with rule-based design decisions. For each section, an LLM generates a slide title, bullet points, a takeaway message, and a visual description that is used to create an illustration. Rule-based matching then selects layouts, themes, and styles to produce the final slides. A vision LLM extracts slide text as Markdown.
We evaluate multiple ASR and spoken language models (SLMs). The best model achieves an average micro-WER of 10.23\% and an average micro-CER of 6.48\% on audio-only inputs. English yields the lowest error rates, followed by German and Dutch, while performance declines across lower-resource languages. Although audio-only baselines are strong, multimodal zero-shot prompting of omni models remains challenging. The aligned slide, text, and audio data show a strong potential to improve recognition through cross-modal context.
\end{abstract}

\begin{IEEEkeywords}
Automatic speech recognition, multimodal speech recognition, contextual biasing, multilingual speech processing, large language models, parameter-efficient fine-tuning, multimodal alignment, Spoken Wikipedia Corpus.
\end{IEEEkeywords}

\section{Introduction}

Automatic speech recognition (ASR) is increasingly integrated with large language models (LLMs), giving rise to spoken language models (SLMs) that jointly process speech and text context~\cite{Arora-2025}. Recent work has explored contextual and multimodal capabilities in tasks such as speech question answering, instruction following, and speech-to-slide alignment~\cite{Peng-2021, anderer24_interspeech, Ranzenberger-2025}.

A common application is the transcription of presentations, including lectures, conference talks, webinars, and business presentations. In these scenarios, speech is often accompanied by slides that provide valuable context such as names, technical terms, abbreviations, formulas, and domain-specific vocabulary. Human listeners naturally exploit this information, and ASR systems should ideally do the same.

Previous work has investigated slide-enhanced transcription of conference talks~\cite{Sinhamahapatra-2025}, but existing datasets are small, focused mainly on English, and often not publicly available. As a result, research on reproducible, multilingual slide-conditioned ASR remains limited.

To address this gap, we introduce the Spoken Wikipedia Presentation Corpus (SWPC)\footnote{\url{https://huggingface.co/datasets/th-nuernberg/swpc}}, a multilingual benchmark built from the Spoken Wikipedia Corpora (SWC)~\cite{Koehn-2016, Baumann-2017} and enriched with synthetic presentation slides. Each sample contains audio, a transcript, and slide-based context in English, German, or Dutch. Slides are generated from Wikipedia article sections using a hybrid pipeline that combines LLM-based planning with deterministic rendering. The corpus also provides generated illustrations and slide Markdown extracted via a vision-language OCR model.

SWPC supports multiple forms of contextual biasing: models can consume slide Markdown as text context or use slide images directly. This enables comparisons across conventional ASR systems, multimodal omni models, and unified LLMs.

Our contributions are threefold: (1) we introduce SWPC, an open multilingual benchmark for slide-conditioned ASR; (2) we release audio, transcripts, slide images, slide Markdown, illustration prompts, and illustrations; and (3) we benchmark ASR and SLM baselines using audio-only input as well as multimodal prompting with Markdown and image context.

\begin{figure}[t]
\centerline{\includegraphics[width=0.92\linewidth]{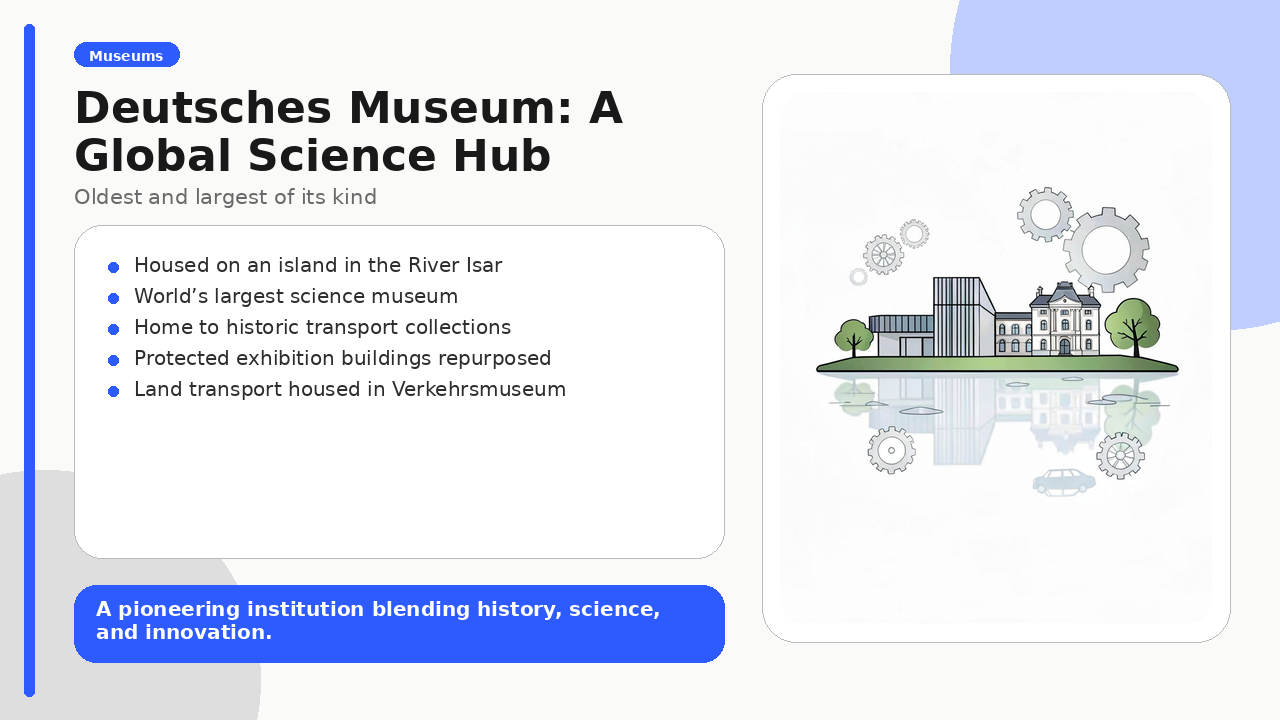}}
\vspace{-3mm}
\caption{Example of a rendered SWPC slide with title, bullet points, takeaway message, and generated illustrative image on the right.}
\label{fig:slide}
\vspace{-3mm}
\end{figure}

\section{Related Work}

While AVSR has traditionally relied on visual speech cues such as lip movements, recent work has explored presentation slides as an additional context source. SlideSpeech~\cite{Wang-2024} introduces a large-scale corpus of conference videos with synchronized speech and slides, showing that slide text improves ASR. SlideAVSR~\cite{wang2024slideavsr} similarly leverages slide text in scientific presentations and proposes DocWhisper to improve recognition of technical terminology.

Most closely related is the benchmark of \cite{Sinhamahapatra-2025}, which demonstrates that slide context reduces word error rates in presentation transcription. In contrast, SWPC provides a multilingual benchmark (English, German, and Dutch) based on Spoken Wikipedia and evaluates both slide text and slide images as contextual information for ASR, SLMs, and multimodal LLMs.

A related research direction investigates aligning presentation videos with slide decks. MaViLS~\cite{anderer24_interspeech} introduces a benchmark and multimodal alignment approach combining speech, OCR, and visual features, finding OCR particularly effective. \cite{Ranzenberger-2025} further demonstrate the value of visually grounded slide representations for speech-to-slide alignment in academic lectures using vision LLMs.

Recent work has advanced automatic presentation generation through agentic and iterative refinement frameworks, including PPTAgent~\cite{Zheng-2025}, EvoPresent~\cite{Liu-2025}, and DeepPresenter~\cite{Zheng-2026}. These systems focus on improving slide quality, design, and coherence through editing, feedback, and visual-grounded revision. In contrast, SWPC does not target presentation generation itself; instead, it uses generated slides as controlled multimodal context for benchmarking slide-conditioned transcription.

\section{Dataset}

\input{figures/dataset_workflow}

We present the Spoken Wikipedia Presentation Corpus (SWPC), a presentation-centric extension of the Spoken Wikipedia Corpora (SWC)~\cite{Koehn-2016, Baumann-2017}. SWPC pairs aligned speech segments with generated presentation slides. Each sample contains an utterance-level audio segment, transcript, rendered slide image, OCR-extracted slide Markdown, illustration image, and illustration prompt. The corpus covers English, German, and Dutch and is released as an open benchmark for multimodal contextual ASR. Figure~\ref{fig:swpc_workflow} outlines the creation pipeline. Figure~\ref{fig:slide} shows an example slide for the utterance:
\begin{tcolorbox}[colframe=black!60, colback=gray!10, boxrule=0.5pt, width=\columnwidth, boxsep=1pt, left=3pt, right=3pt, top=3pt, bottom=3pt]
\textit{``Three redundant exhibition buildings which are under a protection order were converted to house the Verkehrsmuseum, which houses the land transport collections of the Deutsches Museum.''}
\end{tcolorbox}

Utterance-level alignments are obtained from SWC. Segment boundaries are defined by the first and last aligned word timestamps, and audio is extracted as 16~kHz mono WAV. Each utterance is linked to its nearest Wikipedia section heading. Wikipedia articles are converted to Markdown with non-content elements removed. Section text is extracted and, when necessary, subdivided by an LLM at sentence boundaries. This yields coherent text units suitable for slide generation.

SWC reader metadata is often inconsistent. We therefore derive stable speaker identities using an embedding-based registry. Speaker embeddings are extracted with \texttt{speechbrain/spkrec-ecapa-voxceleb}~\cite{Desplanques-2020, speechbrain} after removing low-energy audio. Multiple chunk embeddings are averaged per candidate speaker and matched against persistent speaker centroids. A cosine-similarity threshold of 0.72 determines whether an existing speaker is reused or a new identifier is created.

\begin{table}[t]
\caption{Rule-based mapping from category to slide theme and layout.}
\label{tab:category_mapping}
\centering
\small
\setlength{\tabcolsep}{2.8pt}
\renewcommand{\arraystretch}{0.95}
\begin{tabular}{lll}
\toprule
\textbf{Category} & \textbf{Theme} & \textbf{Layout} \\
\midrule
Literary analysis & Literature & Editorial cards \\
Historical context & History & Timeline visual \\
Person biography & Biography & Portrait focus \\
Place geography & Geography & Map focus \\
Science technical & Science & Concept system \\
Arts culture & Arts & Gallery focus \\
General encyclopedic & Default & Balanced split \\
\bottomrule
\end{tabular}%
\vspace{-3mm}
\end{table}

One slide is generated for each text-segment subsection and assigned to all corresponding utterances. Articles are first mapped to one of seven layout categories (Table~\ref{tab:category_mapping}), which define slide structure and visual style. Structured slide content is generated with \texttt{mistralai/Mistral-Medium-3.5-128B}. The model produces a title, subtitle, bullet points, takeaway statement, illustration description, and speaker notes in constrained JSON format. Slides are rendered deterministically as 720p PNG images containing the generated text and illustration.

The illustrations are generated with a diffusion model \texttt{black-forest-labs/FLUX.2-klein-4B} from the selected visual style and generated illustration brief. Prompts explicitly prohibit textual content to minimize image-generated text artifacts. Final slide images are processed with \texttt{lightonai/LightOnOCR-2-1B}~\cite{Taghadouini-2026} to recover visible slide text as Markdown. OCR output serves both as textual slide context and as a safeguard against unintended textual content in illustrations.

To minimize potential alignment errors, we carry out a preliminary check of the utterances using \texttt{whisper-s2t} and \texttt{systran/faster-whisper-large-v3}~\cite{Radford-2023}. Utterances with more than four reference words and WER above 50\% are discarded. Transcripts containing ``wikipedia'' are also removed, resulting in 317 discarded utterances in total. Final Parquet files are validated to ensure speaker-disjoint splits and integrity of embedded audio and image data.

SWPC is distributed as Parquet. Each row represents a single aligned utterance and includes metadata, transcript, audio, slide image, slide Markdown, illustration image, and illustration prompt. Data are partitioned into speaker-disjoint train, development, and test sets with target ratios of 80/10/10 per language. Cross-lingual speaker occurrences are assigned to a single split. Split assignment is performed using speaker-level duration balancing while preserving speaker exclusivity.

\begin{table}[t]
\caption{SWPC dataset statistics for train, development, test, and total splits. We report utterance, article, text-segment, unique slide, speaker counts, and aligned audio duration.}
\vspace{-2mm}
\label{tab:swpc_dataset_stats}
\centering
\setlength{\tabcolsep}{3.0pt}
\renewcommand{\arraystretch}{0.95}
\resizebox{\columnwidth}{!}{%
\begin{tabular}{llrrrrrr}
\toprule
\textbf{Split} & \textbf{Lang.} & \textbf{Utterances} & \textbf{Articles} & \textbf{Text Seg.} & \textbf{Slides} & \textbf{Speakers} & \textbf{Duration} \\
\midrule
Train & EN & 69,672 & 4,950 & 13,636 & 14,138 & 363 & 149:09:36 \\
Train & DE & 86,354 & 6,067 & 14,823 & 15,684 & 281 & 213:38:15 \\
Train & NL & 36,842 & 3,890 & 7,094 & 7,322 & 57 & 76:12:42 \\
Train & Total & 192,868 & 14,907 & 35,553 & 37,144 & 701 & 439:00:33 \\
\midrule
Dev & EN & 7,531 & 597 & 1,460 & 1,526 & 46 & 19:01:12 \\
Dev & DE & 9,959 & 712 & 1,730 & 1,823 & 35 & 27:15:15 \\
Dev & NL & 4,902 & 616 & 945 & 962 & 15 & 09:39:56 \\
Dev & Total & 22,392 & 1,925 & 4,135 & 4,311 & 96 & 55:56:23 \\
\midrule
Test & EN & 8,818 & 706 & 1,724 & 1,806 & 45 & 21:13:25 \\
Test & DE & 10,883 & 849 & 1,856 & 1,972 & 35 & 27:47:30 \\
Test & NL & 5,644 & 844 & 1,211 & 1,243 & 79 & 09:32:30 \\
Test & Total & 25,345 & 2,399 & 4,791 & 5,021 & 159 & 58:33:25 \\
\midrule
Total & EN & 86,021 & 6,253 & 16,820 & 17,470 & 454 & 189:24:13 \\
Total & DE & 107,196 & 7,628 & 18,409 & 19,479 & 351 & 268:41:00 \\
Total & NL & 47,388 & 5,350 & 9,250 & 9,527 & 151 & 95:25:08 \\
Total & Total & 240,605 & 19,231 & 44,479 & 46,476 & 956 & 553:30:21 \\
\bottomrule
\end{tabular}%
}
\vspace{-2mm}
\end{table}

Table~\ref{tab:swpc_dataset_stats} summarizes corpus statistics. Unique slide counts are computed from hashes of rendered slide images because multiple utterances may share the same slide. SWPC comprises 240,605 utterances from 19,231 articles, yielding 44,479 text segments and 46,476 unique slides. The corpus contains 553.5~h of aligned speech across 956 speakers in three languages, providing a large-scale benchmark for slide-conditioned and multimodal contextual ASR.

\section{Experiments and Results}

\begin{table}[th]
  \caption{SWPC results for baseline ASR and multimodal speech models, sorted by parameter count. Micro/macro WER and CER (\%) are reported for English, German, Dutch, and averaged.}
  \vspace{-2mm}
  \label{tab:swpc_baseline_results}
  \centering
  \scriptsize
  \setlength{\tabcolsep}{2.5pt}
  \renewcommand{\arraystretch}{0.95}
  \resizebox{\columnwidth}{!}{%
  \begin{tabular}{llcccc}
    \toprule
    \textbf{Model} & \textbf{Lang.} & \textbf{WER$_\mu$} & \textbf{WER$_M$} & \textbf{CER$_\mu$} & \textbf{CER$_M$} \\
    \midrule
    \multirow{4}{*}{\texttt{whisper-large-v3}} & EN & 8.14 & \underline{9.78} & 5.90 & \underline{7.04} \\
      & DE & 10.60 & 12.12 & 6.25 & 7.58 \\
      & NL & \underline{17.21} & \underline{23.36} & 10.27 & 13.57 \\
      & Avg. & \underline{10.77} & \underline{13.81} & \underline{6.81} & \underline{8.73} \\
    \midrule
    \multirow{4}{*}{\texttt{canary-1b-v2}} & EN & 9.85 & 12.82 & 6.83 & 8.86 \\
      & DE & 11.09 & 13.61 & 6.38 & 8.26 \\
      & NL & 18.39 & 24.38 & \underline{10.22} & \underline{13.07} \\
      & Avg. & 11.87 & 15.74 & 7.19 & 9.54 \\
    \midrule
    \multirow{4}{*}{\texttt{granite-speech-4.1-2b}} & EN & 7.99 & 10.18 & \underline{5.47} & 7.09 \\
      & DE & 12.92 & 14.34 & 8.36 & 9.41 \\
      & NL & 126.10 & 140.14 & 74.66 & 79.05 \\
      & Avg. & 30.67 & 40.91 & 18.64 & 24.11 \\
    \midrule
    \multirow{4}{*}{\texttt{cohere-transcribe-03-2026}} & EN & \underline{7.76} & 9.80 & 5.58 & 7.06 \\
      & DE & \underline{10.03} & \underline{11.77} & \underline{5.92} & \underline{7.46} \\
      & NL & \textbf{16.35} & \textbf{21.15} & \textbf{9.93} & \textbf{12.36} \\
      & Avg. & \textbf{10.23} & \textbf{13.17} & \textbf{6.48} & \textbf{8.41} \\
    \midrule
    \multirow{4}{*}{\texttt{Phi-4-multimodal-instruct}} & EN & 9.44 & 12.22 & 6.18 & 7.88 \\
      & DE & \textbf{9.99} & \textbf{11.68} & \textbf{5.59} & \textbf{7.10} \\
      & NL & 122.68 & 128.94 & 68.86 & 71.33 \\
      & Avg. & 29.39 & 37.98 & 16.60 & 21.67 \\
    \midrule
    \multirow{4}{*}{\texttt{Qwen2.5-Omni-7B}} & EN & 8.21 & 10.25 & 5.73 & 7.77 \\
      & DE & 11.56 & 12.91 & 6.57 & 7.66 \\
      & NL & 21.82 & 27.04 & 12.96 & 15.27 \\
      & Avg. & 12.01 & 15.13 & 7.36 & 9.39 \\
    \midrule
    \multirow{4}{*}{\texttt{Voxtral-Small-24B-2507}} & EN & \textbf{6.32} & \textbf{8.69} & \textbf{4.61} & \textbf{6.29} \\
      & DE & 14.37 & 18.70 & 8.88 & 13.07 \\
      & NL & 32.42 & 58.95 & 22.87 & 46.46 \\
      & Avg. & 14.32 & 24.18 & 9.74 & 18.14 \\
    \bottomrule
  \end{tabular}%
  }
  \vspace{-5mm}
\end{table}

We evaluate SWPC in three settings. First, audio-only baselines assess ASR systems, spoken language models (SLMs), and multimodal omni models using only the utterance audio. Second, we test zero-shot contextual prompting with the OCR-derived slide Markdown provided by SWPC:
\begin{lstlisting}[style=promptstyle]
Transcribe the given speech referring to the following
words included in markdown wherever needed:
<markdown>
\end{lstlisting}
Third, multimodal models are prompted with the rendered slide image instead of extracted text:
\begin{lstlisting}[style=promptstyle]
Transcribe the given speech referring to the words included
in the image wherever needed.
\end{lstlisting}
These settings measure whether textual or visual slide context improves transcription without task-specific fine-tuning or, for images, an external OCR step.

We report WER and CER using \texttt{jiwer}. Micro scores ($_\mu$) are computed over all utterances per language, while macro scores ($_M$) average per-utterance errors. The Average row aggregates English, German, and Dutch. References and hypotheses are normalized before scoring: English follows the Whisper evaluation setup, while German and Dutch use default normalizers with spoken numbers mapped to digits. German vowel variants and model-specific artifacts, such as \texttt{\$SILENCE}, \texttt{Unk.}, and recurring assistant-style boilerplate, are removed. Table~\ref{tab:swpc_baseline_results} summarizes the benchmark results for all models evaluated, sorted by parameter count. Overall, \texttt{cohere-transcribe-03-2026} achieves the best average performance and is also the top-performing model for Dutch. \texttt{Voxtral-Small-24B-2507} performs best on English, while \texttt{Phi-4-multimodal-instruct} achieves the best German results. Whisper ranks second overall in average performance.
Zero-shot prompting with full slide Markdown or slide images performs poorly for both evaluated multimodal speech models, suggesting that the models are not trained to effectively contextualize speech using text or images.
For all languages, \texttt{Phi-4-multimodal-instruct} and \texttt{Qwen2.5-Omni-7B} exceed 90\% WER/CER, making the outputs impractical. Unlike~\cite{Sinhamahapatra-2025}, who prompt with extracted special terms, SWPC provides complete slide Markdown context. The models frequently perform implicit OCR, copy slide text, or add explanatory content instead of returning only the transcript, causing many insertion errors. These results suggest that effective use of SWPC requires task-specific adaptation or fine-tuning.

\section{Conclusion}
We introduced SWPC, a multilingual benchmark that extends the Spoken Wikipedia Corpora with aligned speech, transcripts, rendered slides, OCR-derived Markdown, illustrations, and prompts for reproducible slide-conditioned ASR research.  
Audio-only baselines achieve strong results, while zero-shot prompting with full slide Markdown or slide images remains ineffective for the benchmarked omni models due to insertion errors from copied slide text, implicit OCR, and non-transcription outputs.
These results establish SWPC as a strong benchmark for future work on multimodal LoRA adapters, parameter-efficient adaptation, and full fine-tuning of ASR models using slide text and visual context.

\section*{AI-Generated Content Disclosure}

The authors used Microsoft Copilot to assist with rewriting, coding, and language polishing. Final manuscript content and code were reviewed, verified, and approved by the authors.

\bibliographystyle{IEEEtran}
\bibliography{mybib}

\end{document}

%% file: figures/dataset_workflow.tex
\begin{figure*}[t]
\centering
\resizebox{0.98\textwidth}{!}{%
\begin{tikzpicture}[
    font=\sffamily\footnotesize,
    >=Latex,
    stage/.style={
        rounded corners=2mm,
        draw=black!65,
        line width=0.5pt,
        align=center,
        text width=25mm,
        minimum height=18mm,
        inner sep=4pt,
        fill=gray!4
    },
    source/.style={
        stage,
        fill=blue!6,
        draw=blue!55!black
    },
    output/.style={
        stage,
        fill=green!7,
        draw=green!45!black,
        text width=28mm
    },
    arrow/.style={
        ->,
        line width=0.8pt,
        draw=blue!65!black,
        shorten >=1pt,
        shorten <=1pt
    }
]

\node[source] (src) at (0,0) {
    \textbf{1. SWC sources}\\
    audio alignments\\
    Wikipedia articles
};

\node[stage] (prep) at (3.0,0) {
    \textbf{2. Preprocessing}\\
    cut utterances\\
    identify speakers\\
    segment text
};

\node[stage] (map) at (6.0,0) {
    \textbf{3. Mapping}\\
    link utterances\\
    to article\\
    subsections
};

\node[stage] (slides) at (9.0,0) {
    \textbf{4. Slide generation}\\
    plan slide\\
    create illustration\\
    render image
};

\node[stage] (ocr) at (12.0,0) {
    \textbf{5. OCR}\\
    extract Markdown\\
};

\node[stage] (qc) at (15.0,0) {
    \textbf{6. Filtering}\\
    remove outliers\\
    validate files\\
    split speakers
};

\node[output] (out) at (18.1,0) {
    \textbf{7. SWPC dataset}\\
    audio, transcript\\
    slide image/text\\
    illustration\\
    speaker splits
};

\draw[arrow] (src) -- (prep);
\draw[arrow] (prep) -- (map);
\draw[arrow] (map) -- (slides);
\draw[arrow] (slides) -- (ocr);
\draw[arrow] (ocr) -- (qc);
\draw[arrow] (qc) -- (out);

\end{tikzpicture}%
}
\vspace{-2mm}
\caption{High-level overview of the SWPC data creation workflow. Spoken Wikipedia audio and article text are processed into aligned utterance--section mappings, which are used to generate slide images, extract slide text, apply quality control, and assemble the final multimodal dataset.}
\label{fig:swpc_workflow}
\vspace{-6mm}
\end{figure*}
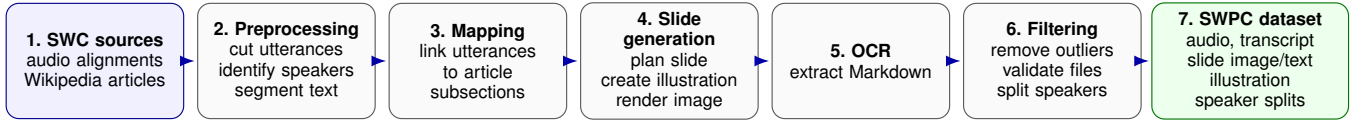